\documentclass[aps,pr,twocolumn,superscriptaddress,groupedaddress,nofootinbib]{revtex4}  

\usepackage{graphicx}  
\usepackage{dcolumn}   
\usepackage{bm}        
\usepackage{amssymb}  
\usepackage{amssymb}
\usepackage{amsxtra}
\usepackage{euscript} 
\usepackage{amsthm}
\usepackage{color}
\usepackage{ulem}

\usepackage{bbold}
\usepackage{pifont}
\usepackage{dsfont}
\usepackage{MnSymbol}

\usepackage{color}
\usepackage{listings}

\begin{document}

\title{Infrared propagators of Yang-Mills-Chern-Simons theories in linear covariant gauges}

\author{Luigi C. Ferreira}
\email{luigi_carvalho@id.uff.br}
\affiliation{Instituto de F\'isica, Universidade Federal Fluminese, Av. Litor\^anea s/n, 24210-346, Niter\'oi, RJ, Brazil}

\author{Diego R. Granado}
\email{diegorochagranado@duytan.edu.vn}
\affiliation{Institute of Research and Development, Duy Tan University, Da Nang 550000, Viet Nam}
\affiliation{Faculty of Natural Sciences, Duy Tan University, Da Nang 550000, Viet Nam}

\author{Igor F. Justo}
\email{ijusto@id.uff.br}
\affiliation{Instituto de F\'isica, Universidade Federal Fluminese, Av. Litor\^anea s/n, 24210-346, Niter\'oi, RJ, Brazil}

\author{Antonio D. Pereira}
\email{A.Duarte@science.ru.nl}
\affiliation{Institute for Mathematics, Astrophysics and Particle Physics (IMAPP), Radboud University, Heyendaalseweg 135, 6525 AJ Nijmegen, The Netherlands}
\affiliation{Instituto de F\'isica, Universidade Federal Fluminese, Av. Litor\^anea s/n, 24210-346, Niter\'oi, RJ, Brazil}

\begin{abstract}
Recent works have explored non-perturbative effects due to the existence of (infinitesimal) Gribov copies in Yang-Mills-Chern-Simons theories in three Euclidean dimensions. In particular, the removal of such copies modify the gauge field propagator by a self-consistent dynamically  generated mass parameter, the Gribov parameter. Due to the interplay with the topological mass introduced by the Chern-Simons term, the propagator features a non-trivial set of phases with poles of different nature, leading to the possible interpretation of a confinfing to deconfining phase transition. Inhere, we restore the BRST symmetry which is softly broken by the elimination of gauge copies and provide a BRST-invariant discussion of such a transition. In order to make clear all physical statements, we deal with linear covariant gauges which contain a gauge parameter and therefore allow for an explicit check of gauge parameter independence of physical results. We also discuss the generation of condensates due to the infrared relevance of infinitesimal Gribov copies.
\end{abstract}

\maketitle

\section{Introduction}

\subsection{Gauge-fixing pure Yang-Mills theories in the infrared}

The quantization of Yang-Mills theories\footnote{We restrict all the statements of this paper to Yang-Mills theories formulated in Euclidean space.} in a continuum framework typically requires a gauge-fixing procedure. In a path-integral quantization, this is achieved by the celebrated Faddeev-Popov procedure \cite{Faddeev:1967fc}. However, since the seminal work by Gribov \cite{Gribov:1977wm} followed by Singer's mathematical formalization and generalization \cite{Singer:1978dk}, it became clear that gauge fixings, which are continuous in field space, do not select a single representative per gauge orbit, in general. The standard situation is that the gauge fixing section crosses a gauge orbit several times and many configurations, which obey the gauge condition and belong to the same gauge orbit (i.e., that can be connected by a gauge transformation), are picked up by the gauge fixing procedure. Such spurious configurations are known as Gribov copies and, in the Faddeev-Popov method, they are assumed to not exist. Albeit there is no objection about their existence, one might wonder why perturbative calculations tacitly ignore such copies and seem to provide results which agree very well with high-energy experiments. Despite the possibility that such gauge copies might have their effect suppressed by some miraculous cancellation, it is possible to show that perturbative calculations performed around the perturbative vacuum, where the gauge field is a vanishing field configuration, i.e., $A^{a}_\mu = 0$, gauge copies seem to not be generated. Geometrically, this has the simple interpretation that for such a calculation, it suffices to trace a local section across the orbits, a necessity that can be achieved. However, the larger the field configuration gets, one can easily argue that Gribov copies can be generated, see, e.g., \cite{Gribov:1977wm,Sobreiro:2005ec}. Hence, it can be expected that towards low energies (or growing coupling constant), the Faddeev-Popov procedure must be improved and replaced by another method which takes into account the existence of Gribov copies. So far, this has been achieved just for the infinitesimal copies, i.e., those generated by infinitesimal gauge transformations. In fact, those copies are generated by zero-modes of the Faddeev-Popov operator. As proposed in \cite{Gribov:1977wm} and \cite{Zwanziger:1989mf}, a possible way to deal with infinitesimal copies is to restrict the functional integral to a region where the spectrum of the Faddeev-Popov operator does not have any zero-modes. In the Landau gauge, the Faddeev-Popov operator is Hermitian and it is possible to define a region where such an operator is positive. This region is the so-called Gribov region $\Omega$ and it features very important geometrical properties \cite{DellAntonio:1991mms}. A key property is that every gauge orbit cross such a region at least once and, therefore, every configuration outside of it has a copy that lies inside $\Omega$. Consequently, restricting the path integral to $\Omega$ takes into account all unequivalent field configurations and eliminate the infinitesimal Gribov copies. Nevertheless, it should be clear that $\Omega$ is not free of copies, due to the presence of those generated by finite gauge transformations. A region which is truly free of gauge copies is the so-called fundamental modular region (FMR) \cite{vanBaal:1991zw}. Up to now, there is no systematic method that can restrict the functional integration to the FMR. Thus, restricting to $\Omega$ corresponds to a first step towards the removal of all Gribov copies from the gauge-fixing procedure. 

One practical way of imposing the Gribov restriction to the path integral, in the Landau gauge, was proposed in \cite{Gribov:1977wm} to leading order, and in \cite{Zwanziger:1989mf} at all orders, but using a different implementation. The equivalence of the methods was proved much later in \cite{Capri:2012wx}. For a review, we refer to \cite{Vandersickel:2012tz}.  Effectively, the restriction to $\Omega$ is achieved by the introduction of the so-called horizon function $H(A)$ to the standard Yang-Mills action together with the Faddeev-Popov gauge-fixing term. Hence, the action that enters the Boltzmann weight and implements the restriction to $\Omega$, in $d$ dimensions and for the gauge group $SU(N)$, is\footnote{We adopt the short-hand notation $\int \mathrm{d}^dx = \int_{x^d}$. In three dimensions - which is the focus of this paper, we simply write $\int \mathrm{d}^3x = \int_x$.}
\begin{equation}
S^{\mathrm{nl}}_{\mathrm{GZ}} = S_{\mathrm{YM}} + S_{\mathrm{FP}} + S_{\mathrm{H}}\,,
\label{Intro1}
\end{equation}
with
\begin{equation}
S_{\mathrm{YM}} = \frac{1}{4}\int_{x^d} F^{a}_{\mu\nu}F^{a}_{\mu\nu}\,.
\label{Intro2}
\end{equation}
The Faddeev-Popov action in the Landau gauge is
\begin{equation}
S_{\mathrm{FP}} = \int_{x^d}\left(b^a\partial_\mu A^a_\mu + \bar{c}^a\partial_\mu D^{ab}_{\mu}c^b\right)\,,
\label{Intro3}
\end{equation}
and
\begin{eqnarray}
S_{\mathrm{H}} &=& g^2\gamma^4\int_{x^d}f^{abc}A^{b}_\mu\left[-\left(\partial_\alpha D_\alpha\right)^{-1}\right]^{ad}f^{dec}A^{e}_\mu\nonumber\\
&=& \gamma^4 H(A)\,.
\label{Intro4}
\end{eqnarray}
The field strength is defined as $F^{a}_{\mu\nu} = \partial_\mu A^a_\nu - \partial_\mu A^a_\nu + gf^{abc}A^{b}_\mu A^{c}_\nu$, the covariant derivative in the adjoint representation of the gauge group is $D^{ab}_\mu = \delta^{ab}\partial_\mu - gf^{abc}A^c_\mu$, with $g$ being the coupling constant, and $f^{abc}$ are the structure constants of $SU(N)$. The fields $b^a$, $\bar{c}^a$, and $c^a$ are, respectively, the Lagrange multiplier that imposes the gauge condition, and the Faddeev-Popov ghosts. The parameter $\gamma$ is known as the Gribov parameter and is determined by a gap equation, namely, 
\begin{equation}
\langle H(A)\rangle = dV(N^2-1)\,,
\label{Intro5}
\end{equation}
with $V$ being the (regularized) volume of spacetime.  As it is clearly seen in \eqref{Intro4}, the horizon function is non-local. However, as introduced in \cite{Zwanziger:1989mf}, the action \eqref{Intro1} can be cast in a local form by the use of a suitable set of auxiliary fields. In particular, the so-called Gribov-Zwanziger (GZ) action is written as
\begin{eqnarray}
S_{\mathrm{GZ}} &=& S_{\mathrm{YM}} + S_{\mathrm{FP}} + s\int_{x^d}\bar{\omega}^{ac}_\mu\partial_\alpha D^{ab}_\alpha \varphi^{bc}_\mu\nonumber\\
&+&\gamma^2\int_{x^d}\,gf^{abc}A^{a}_\mu (\bar{\varphi}+\varphi)^{bc}_\mu\,,
\label{Intro6}
\end{eqnarray}
where $s$ stands for the nilpotent BRST operator, which acts on the complete set of fields as
\begin{align}
sA^{a}_{\mu}&=-D^{ab}_{\mu}c^b\,,     &&sc^a=\frac{g}{2} f^{abc}c^b c^c\,, \nonumber\\
s\bar{c}^a&=b^{a}\,,     &&sb^{a}=0\,, \nonumber\\
s\bar{\omega}^{ab}_\mu &=\bar{\varphi}^{ab}_{\mu} \,,     &&s\bar{\varphi}^{ab}_{\mu} = 0\,,\nonumber\\
s\varphi^{ab}_\mu &=\omega^{ab}_{\mu} \,,     &&s\omega^{ab}_{\mu} = 0\,.
\label{Intro7}
\end{align}
The fields $(\bar{\varphi},\varphi)^{ab}_\mu$ are commuting ones, while $(\bar{\omega},\omega)^{ab}_\mu$ are anti-commuting. An important consequence concerning the construction of the GZ action in the Landau gauge is that it breaks BRST invariance, i.e.,
\begin{equation}
sS_{\mathrm{GZ}} = \gamma^2 \int_{x^d}\,gf^{abc}\left[-(D^{ad}_\mu c^d)(\bar{\varphi}+\varphi)^{bc}_{\mu}+A^{a}_\mu\omega^{bc}_{\mu}\right]\,.
\label{Intro8}
\end{equation}
The breaking in eq.\eqref{Intro8} is explicit but soft in the sense that it is proportional to the Gribov parameter and therefore vanishes in the deep ultraviolet regime. 

Since the standard BRST invariance of the Faddeev-Popov action is a direct outcome of the Faddeev-Popov quantization, it can be expected that the BRST symmetry will be deformed at low energies if the gauge-fixing procedure is affected in a such energy regime. This was an open issue for several years and many works were done in order to better understand the fate of BRST symmetry in the infrared, see, e.g., \cite{Maggiore:1993wq,Baulieu:2008fy,Dudal:2009xh,Sorella:2009vt,Sorella:2010it,Capri:2010hb,Dudal:2012sb,Capri:2014bsa,Cucchieri:2014via,Schaden:2014bea,Schaden:2015uua,Lavrov:2011wb,Lavrov:2013boa,Moshin:2014xka,Moshin:2015gsa,Serreau:2012cg,Serreau:2013ila,Serreau:2015yna,Pereira:2013aza,Pereira:2014apa}. In \cite{Capri:2015ixa}, it was realized that BRST invariance can be achieved by a suitable modification of the horizon function which corresponds to a dressing of the gauge field by a gauge-invariant variable $A^{h,a}_\mu$. This has led to the proposal of a non-perturbative BRST quantization that takes into account the existence of (infinitesimal) Gribov copies in the gauge-fixing procedure, \cite{Capri:2015nzw,Pereira:2016fpn,Capri:2016aqq,Capri:2016gut,Capri:2017bfd,Capri:2018ijg}. In the present work, we will apply such a quantization procedure to Yang-Mills-Chern-Simons (YMCS) theories quantized in the linear covariant gauges and inspect the consequences to the spectrum of the theory.

\subsection{YMCS theories and infinitesimal Gribov copies}

The YMCS action is defined in three Euclidean dimensions as follows,
\begin{equation}
S_{\mathrm{YMCS}} = S_{\mathrm{YM}}+S_{\mathrm{CS}}\,,
\label{YMCSGP1}
\end{equation}
where the Chern-Simons action $S_{\mathrm{CS}}$ being
\begin{equation}
S_{\mathrm{CS}} = -i M \int_x \epsilon_{\mu\rho\nu}\left(\frac{1}{2}A^a_\mu\partial_\rho A^a_\nu + \frac{g}{3!}f^{abc}A^{a}_{\mu}A^{b}_{\rho}A^{c}_{\nu}\right)\,,
\label{YMCSGP2}
\end{equation}
with $M$ being a mass parameter and $\epsilon_{\mu\rho\nu}$ is the totally anti-symmetric Levi-Civita symbol. Due to the topological nature of the Chern-Simons action, the mass $M$ is also known as topological mass and it provides a mass to the gauge field while being compatible with infinitesimal gauge invariance. In order to achieve the invariance under finite gauge transformations, it is necessary to impose a constraint over $M$, which we do not do it here. Such a system is particularly interesting since the introduction of a mass parameter for the gauge field, which is compatible with gauge invariance, is possible in a local way, \cite{Deser:1981wh,Deser:1982vy}. This consists of a rich arena to study the pole structure of the gauge field propagator in three-dimensional gauge theories, rendering some understanding that could be lifted to the four-dimensional at zero or finite temperature. For the Gribov problem and its resolution, this system is particularly important due to the fact that for a generic value of $M$, gauge invariance is reduced to the subset of infinitesimal gauge transformations. Since, at the present moment, we have the limited understanding of eliminating just infinitesimal copies, this system is well-suited for studying the consequences of removing all Gribov copies from a gauge theory.

As before, the Faddeev-Popov procedure can be applied. Since the gauge condition can be taken as Landau gauge $\partial_\mu A^{a}_\mu = 0$, all the discussion made before regarding the existence and treatment of infinitesimal Gribov copies can be simply imported to the case of YMCS theories. This was investigated in \cite{Canfora:2013zza,Gomez:2015aba} in the Landau gauge and in \cite{Ferreira:2020kqv} in the maximal Abelian gauge. As a difference with respect to pure Yang-Mills theories, the topological mass $M$ enters the gap equation that fixes the Gribov parameter as observed in \cite{Ferreira:2020kqv}. See also \cite{Felix:2021eoq}. In this work we propose the quantization of YMCS theories in linear covariant gauges, i.e., gauges of the form \eqref{YMCSGP3}.
\begin{equation}
\partial_\mu A^a_\mu = \alpha b^a\,.
\label{YMCSGP3}
\end{equation} 
The parameter $\alpha$ is non-negative and the condition \eqref{YMCSGP3} essentially provides a fixed longitudinal part to the gauge field. Moreover, the removal of infinitesimal Gribov copies in the gauge \eqref{YMCSGP3} requires a complete BRST-invariant quantization in order to ensure that physical correlators do not depend on the parameter $\alpha$. The gauge-fixed YMCS action in the linear covariant gauges $S^{\mathrm{LCG}}_{\mathrm{YMCS}}$ is written as
\begin{eqnarray}
S^{\mathrm{LCG}}_{\mathrm{YMCS}} &=& S_{\mathrm{YMCS}}\nonumber\\
&+&\int_x\left(b^a\partial_\mu A^a_\mu - \frac{\alpha}{2}b^a b^a + \bar{c}^a \partial_\mu D^{ab}_{\mu} c^b\right)\,.\nonumber\\
\label{YMCSGP4}
\end{eqnarray}
We emphasize that due to the introduction of the Chern-Simons action with generic topological mass, the gauge fixing introduced in
\eqref{YMCSGP4} is needed to fix just the infinitesimal gauge invariance. Thus, for this system, the Gribov copies that will actually be present are those generated by infinitesimal gauge transformations. Finite copies are not redundancies of the full action due to the Chern-Simons term. Consequently, removing the infinitesimal copies is more than a first step towards a complete gauge fixing, but rather a complete gauge fixing in this system. The tree-level gauge field propagator is,
\begin{eqnarray}
\langle A^{a}_{\mu} (p) A^{b}_{\nu} (-p)\rangle &=& \frac{\delta^{ab}}{p^2+M^2}\Bigg(\EuScript{P}^{\mathrm{T}}_{\mu\nu}(p)+ \frac{M}{p^2}\epsilon_{\mu\rho\nu}p_\rho\Bigg)\nonumber\\
&+&\delta^{ab}\frac{\alpha}{p^2}\frac{p_\mu p_\nu}{p^2}\,,
\label{YMCSGP5}
\end{eqnarray}
where $\EuScript{P}^{\mathrm{T}}_{\mu\nu}(p)$ stands for the transverse projector,
\begin{equation}
\EuScript{P}^{\mathrm{T}}_{\mu\nu}(p)=\delta_{\mu\nu}-\frac{p_\mu p_\nu}{p^2}\,.
\label{YMCSGP6}
\end{equation}
For $\alpha\to 0$, eq.\eqref{YMCSGP5} reduces to the tree-level propagator of the gauge field in YMCS in the Landau gauge, see \cite{Canfora:2013zza}. If the Chern-Simons term is removed by taking $M\to 0$, one recovers the gauge field tree-level propagator in pure Yang-Mills theories in the Landau gauge. It is clear from eq.\eqref{YMCSGP5} that $M$ enters as a mass parameter in the gauge-field propagator. As it was investigated in \cite{Canfora:2013zza,Gomez:2015aba,Ferreira:2020kqv,Felix:2021eoq}, the presence of the Gribov parameter $\gamma$ in the gauge-field propagator has a non-trivial interference with $M$, which changes the pole-structure of the gauge-field propagator and the interpretation of (non-)physical excitations in the spectrum of the theory.

The paper is organized as follows: In Sect.~\ref{Sect:NPBRST} we will introduce the non-perturbative BRST quantization of YMCS theories in the linear covariant gauges, i.e., a local and BRST-invariant action that effectively restricts the path integral to a region free of infinitesimal Gribov copies. We will discuss how the limit $\alpha\to 0$ leads to a BRST-invariant action which renders physical correlators which are equivalent to those obtained in the BRST-soft broken version of the theory. Moreover, we collect the tree level gauge-field propagator. In Sect.~\ref{Sect:Condensates}, a discussion regarding further infrared instabilities due to the elimination of Gribov copies is made. This will give rise to a refined version of the theory in analogy to the pure Yang-Mills case, that we will shortly review. Finally, in Sect.~\ref{Sect:CONCLUSIONS} the main results are collectively discussed, and perspectives for future work concerning the present model is presented.

\section{Non-perturbative BRST quantization of YMCS theories}\label{Sect:NPBRST}

\subsection{The local and BRST-invariant action free of infinitesimal Gribov copies}
Following the non-perturbative BRST quantization introduced in\footnote{See \cite{Sobreiro:2005vn,Capri:2015pja} for earlier attempts to deal with the Gribov problem in the linear covariant gauges.} \cite{Capri:2015ixa,Capri:2015nzw,Pereira:2016fpn,Capri:2018ijg,Capri:2016aqq,Capri:2016gut,Capri:2017bfd}, the local action action which renders a gauge-fixed YMCS theory in the linear covariant gauges, and eliminates infinitesimal Gribov copies in harmony with BRST invariance, which we shall call the Gribov-Zwanziger modification to the YMCS action invariant, is
\begin{eqnarray}
S_{\mathrm{GZ}} &=& S^{\mathrm{LCG}}_{\mathrm{YMCS}} - \int_x\Big(\bar{\varphi}^{ac}_{\mu}\EuScript{M}^{ab}(A^h)\varphi^{bc}_\mu\nonumber\\
&-&\bar{\omega}^{ac}_{\mu}\EuScript{M}^{ab}(A^h)\omega^{bc}_\mu\Big)+\gamma^2\int_{x}\,gf^{abc}A^{h,a}_\mu (\bar{\varphi}+\varphi)^{bc}_\mu\nonumber\\
&+&\int_x \Big(\tau^a \partial_\mu A^{h,a}_\mu-\bar{\eta}^a\EuScript{M}^{ab}(A^h)\eta^b\Big)\,.
\label{NPBRSTYMCS1}
\end{eqnarray} 
The localizing fields $(\bar{\varphi},\varphi,\bar{\omega},\omega)^{ab}_\mu$ are now BRST singlets, $A^{h,a}_\mu$ is a gauge-invariant field which is defined according to
\begin{equation}
A^{h,a}_\mu T^a = h^{\dagger}A_\mu h + \frac{i}{g}h^{\dagger}\partial_\mu h\,,
\label{NPBRSTYMCS2}
\end{equation}
where $T^a$ denotes the generators of the $SU(N)$ gauge group. Moreover,
\begin{equation}
h = \mathrm{e}^{ig\xi^a T^a}\equiv \mathrm{e}^{ig\xi}\,,
\label{NPBRSTYMCS3}
\end{equation}
with $\xi= \xi^a T^a$ being a Stueckelberg-like field. In the Appendix~\ref{Appendix:AhConstruction} we collect the main ingredients for the construction of the dressed gauge-invariant field $A^h_\mu$. The operator $\EuScript{M}^{ab}(A^h)$ is equivalent to the Faddeev-Popov operator where the gauge field $A^a_\mu$ is replaced by the dressed field $A^{h,a}_\mu$, i.e.,
\begin{equation}
\EuScript{M}^{ab}(A^{h}) = - \partial_\mu D^{ab}_\mu (A^h) = \partial_\mu (\delta^{ab}\partial_\mu -gf^{abc}A^{h,c}_\mu)\,.
\label{NPBRSTYMCS4}
\end{equation}
The field $\tau^a$ in eq.\eqref{NPBRSTYMCS1} works as a Lagrange multiplier that imposes the transversality condition over $A^{h,a}_\mu$, a property that is explained in the Appendix~\ref{Appendix:AhConstruction}. Such a constraint, due to the composite nature of $A^h_\mu$, demands the introduction of a Jacobian compensator, which is encoded by the extra ghosts $(\bar{\eta},\eta)^a$. Therefore, the path integral associated with the Gribov-Zwanziger modification to the YMCS action in the linear covariant gauges is expressed as
\begin{equation}
\EuScript{Z} = \int [\EuScript{D}\Phi]_{\mathrm{GZ}}\,\mathrm{e}^{-S_{\mathrm{GZ}}-3V\gamma^4(N^2-1)}\,,
\label{NPBRSTYMCS5}
\end{equation}
with 
\begin{eqnarray}
[\EuScript{D}\Phi] &=& [\EuScript{D}A][\EuScript{D}b][\EuScript{D}\bar{c}][\EuScript{D}c][\EuScript{D}\bar{\varphi}][\EuScript{D}\varphi][\EuScript{D}\bar{\omega}][\EuScript{D}\omega]\nonumber\\
&\times&[\EuScript{D}\xi][\EuScript{D}\tau][\EuScript{D}\bar{\eta}][\EuScript{D}\eta]\,,
\label{NPBRSTYMCS6}
\end{eqnarray}
being the new functional measure for the local theory. The complete set of BRST transformations that leaves the action \eqref{NPBRSTYMCS1} invariant is
\begin{align}
sA^{a}_{\mu}&=-D^{ab}_{\mu}c^b\,,     &&sc^a=\frac{g}{2} f^{abc}c^b c^c\,, \nonumber\\
s\bar{c}^a&=b^{a}\,,     &&sb^{a}=0\,, \nonumber\\
s\varphi^{ab}_{\mu}&=0\,,   &&s\omega^{ab}_{\mu}=0\,, \nonumber\\
s\bar{\omega}^{ab}_{\mu}&=0\,,         &&s\bar{\varphi}^{ab}_{\mu}=0\,,\nonumber \\
s h^{ij}& = -ig c^a (T^a)^{ik} h^{kj}  \;, && sA^{h,a}_\mu =0\,,   \nonumber \\
s\tau^a& =0\,, && s\bar{\eta}^a=0 \,, \nonumber \\
s\eta^a &= 0\,, && s\xi^a = g^{ab}(\xi)c^b\,,
\label{NPBRSTYMCS7}
\end{align}
with
\begin{equation}
g^{ab}(\xi) =  - \delta^{ab} + \frac{g}{2} f^{abc} \xi^c - \frac{g^2}{12} f^{amr} f^{mbq}  \xi^q \xi^r + O(g^3)\,.
\label{NPBRSTYMCS8}
\end{equation}
The transformations \eqref{NPBRSTYMCS7}, together with \eqref{NPBRSTYMCS8}, are generated by a nilpotent operator $s$, i.e., $s^2=0$. Thus, one sees that the local BRST-invariant Gribov-Zwanziger action \eqref{NPBRSTYMCS1} is composed by the standard Faddeev-Popov gauge-fixed and BRST invariant action, i.e., the YMCS action together with the BRST exact gauge-fixing term. On top of that the BRST invariant contributions are added, which effectively implement the elimination of infinitesimal Gribov copies. Those terms are not BRST exact, but are BRST invariant. Integrating out the auxiliary fields $(\bar{\varphi},\varphi,\bar{\omega},\omega,\xi,\tau,\bar{\eta},\eta)$, one obtain the non-local horizon function $H(A^h)$ written as
\begin{eqnarray}
S^{\mathrm{LCG}}_{\mathrm{H}} &=& g^2\gamma^4\int_{x}f^{abc}A^{h,b}_\mu\left[\EuScript{M}^{-1}(A^h)\right]^{ad}f^{dec}A^{h,e}_\mu\nonumber\\
&=& \gamma^4 H(A^h)\,.
\label{NPBRSTYMCS9}
\end{eqnarray}
In eq.\eqref{NPBRSTYMCS9}, $A^{h,a}_\mu$ is a non-local expression of $A^{a}_{\mu}$, which is presented in the Appendix~\ref{Appendix:AhConstruction}. Hence, the horizon function in the linear covariant gauges features two sources of non-localities: The one coming from $A^{h,a}_\mu$ and the other, which is the same as in expression \eqref{Intro4}, namely, due to the inverse of $\mathcal{M}(A^{h}_{\mu})$. The horizon function \eqref{NPBRSTYMCS9} can be viewed as a dressing of \eqref{Intro4}, where the gauge field $A^a_\mu$ is replaced by a gauge-invariant composite operator $A^{h,a}_\mu$. Geometrically, the horizon function \eqref{NPBRSTYMCS9} has the role of restricting the path integral domain to the region $\Omega_h$ which is defined by
\begin{equation}
\Omega_h = \left\{A^a_\mu\,; \,\partial_\mu A^a_\mu = \alpha b^a\,\, |\,\, \partial_\mu D^{ab}_\mu (A^h)>0\right\}\,.
\label{NPBRSTYMCS10}
\end{equation}
The operator $\partial_\mu D^{ab}_\mu (A^h)$ is Hermitian due to the transversality of $A^{h,a}_\mu$, see Appendix~\ref{Appendix:AhConstruction}. Thereby, the partition function \eqref{NPBRSTYMCS5} is equivalent to
\begin{equation}
\EuScript{Z} = \int_{\Omega^h} [\EuScript{D}\Phi]_{\mathrm{YMCS}}\,\mathrm{e}^{-S^{\mathrm{LCG}}_{\mathrm{YMCS}}-3V\gamma^4(N^2-1)}\,,
\label{NPBRSTYMCS11}
\end{equation}
where $ [\EuScript{D}\Phi]_{\mathrm{YMCS}}$ is the measure with the standard field content of gauge-fixed YMCS theories.

One can easily obtain the Gribov-Zwanziger modification to the YMCS theory in the Landau gauge by taking $\alpha\to 0$. However, from eq.\eqref{NPBRSTYMCS1}, one does not recover \eqref{Intro6} immediately. Nonetheless, those actions are connected at the level of the non-local horizon function by a redefinition of the $b^a$ field due to the observation that the dressed field $A^{h,a}_\mu$ is related to the gauge field by the structure,
\begin{equation}
A^{h,a}_\mu = A^a_\mu - \EuScript{R}^{ab}_\mu (A)(\partial_\alpha A^b_\alpha)\,,
\label{NPBRSTYMCS12}
\end{equation}
where $\EuScript{R}^{ab}_\mu (A)$ is a non-local expression of the gauge field, see Appendix~\ref{Appendix:AhConstruction}. The presence of a divergence of the gauge field ensures that, in the Landau gauge, all the non-localities can be absorbed in the $b$-field redefinition. Alternatively, one sees that the dressed field collapses to the gauge field when the gauge condition is applied to eq.\eqref{NPBRSTYMCS12}. In summary, the dressed horizon function can be reduced to the Landau gauge horizon function \eqref{Intro4} due to the transversality of $A^{h,a}_\mu$. For more details we refer the reader to \cite{Capri:2015ixa,Capri:2018ijg}.

The tree-level gauge-field propagator arising from the GZ modification of the YMCS in the linear covariant gauges is
\begin{eqnarray}
\langle A^{a}_\mu (p) A^b_\nu (-p)\rangle &=& \delta^{ab}\Bigg\{\frac{p^2 (p^4+2g^2\gamma^4 N)}{(p^4+2g^2\gamma^4 N)^2 + M^2 p^6}\Bigg[\EuScript{P}^{\mathrm{T}}_{\mu\nu}(p)\nonumber\\
&+&\frac{Mp^2}{p^4+2g^2\gamma^4 N}\epsilon_{\mu\lambda\nu}p_\lambda\Bigg]+\frac{\alpha}{p^2}\frac{p_\mu p_\nu}{p^2}\Bigg\}\,.\nonumber\\
\label{NPBRSTYMCS13}
\end{eqnarray}
As it is clear from eq.\eqref{NPBRSTYMCS13}, the parity-preserving part of the propagator has just the transverse components, i.e., those proportional to $\EuScript{P}^{\mathrm{T}}_{\mu\nu}(p)$ affected by the introduction of the horizon function. The rest of the parity-preserving components, the longitudinal part, remains the same as with $\gamma=0$. Later on, it will be shown that this property is preserved at all orders in perturbation theory. The parity-violating part is directly affected by the elimination of infinitesimal Gribov copies. Comparing equation \eqref{YMCSGP5} to \eqref{NPBRSTYMCS13} shows explicitly that the pole structure of the gauge-field propagator is affected by the presence of the Gribov parameter $\gamma$. As a consistency check, if $\gamma\to 0$, one recovers \eqref{YMCSGP5} and if $M\to 0$, it coincides with the tree-level gauge-field propagator in the linear covariant gauges, see, e.g., \cite{Capri:2015nzw}. Moreover, the parity-preserving and transverse part of \eqref{NPBRSTYMCS13} does not depend on $\alpha$ at the tree level and coincides with the result in \cite{Canfora:2013zza}. Therefore, there is no need to repeat the analysis of the pole structure inhere. However, we should emphasize that when loop corrections are considered, the transverse part of the propagator will receive $\alpha$-dependent contributions. Nevertheless, as discussed in \cite{Capri:2016gut}, the poles are gauge parameter independent. 

The Gribov parameter $\gamma$ is not free, but it is fixed in terms of the initial parameters of the theory by a gap equation. In particular, the Gribov parameter $\gamma$ in eq.\eqref{NPBRSTYMCS1} does not enter as the BRST variation of anything, i.e.,
\begin{equation}
\frac{\partial S_{\mathrm{GZ}}}{\partial \gamma^2} = \int_x gf^{abc}A^{h,a}_{\mu}(\bar{\varphi}+\varphi)^{bc}_\mu \neq s(\ldots)\,,
\label{NPBRSTYMCS14}
\end{equation}
and thereby can enter correlation function of gauge-invariant operators. In a local setting \eqref{NPBRSTYMCS5}, the gap equation is formulated by an extremization of the vacuum energy $\EuScript{E}_v$ of the theory,
\begin{equation}
\mathrm{e}^{-V\EuScript{E}_v} = \int [\EuScript{D}\Phi]_{\mathrm{GZ}}\,\mathrm{e}^{-S_{\mathrm{GZ}}+3V\gamma^4(N^2-1)}\,,
\label{NPBRSTYMCS15}
\end{equation}
at vanishing sources and fields, i.e.,
\begin{equation}
\frac{\partial\EuScript{E}_v}{\partial \gamma^2} = 0\,.
\label{NPBRSTYMCS16}
\end{equation} 
At one-loop order it leads to the gap equation,
\begin{equation}
\frac{2Ng^2}{3}\int\frac{\mathrm{d}^3p}{(2\pi)^3}\frac{p^4 + 2g^2\gamma^4N}{(p^4 + 2g^2\gamma^4N)^2+M^2p^6} = 1\,.
\label{NPBRSTYMCS17}
\end{equation}
Such an equation fixes $\gamma$ as a function of $M$ and $g$. In three dimensions this integral is convergent and can be solved directly. Moreover, at one loop the gap equation equation does not depend on the gauge parameter $\alpha$. This is completely expected since the gap equation in the non-local form is written as
\begin{equation}
\langle H(A^h)\rangle = 3V(N^2-1)\,,
\label{NPBRSTYMCS18}
\end{equation}
which is manifestly gauge invariant and independent of the gauge parameter $\alpha$. Moreover, since $\gamma$ can enter gauge-invariant correlation functions, it carries a physical character and should not be $\alpha$-dependent.  

\subsection{Exactness of the longitudinal part of the gauge field propagator}

In the above discussions about the tree-level gauge-field propagator, we have seen that the longitudinal parity-preserving sector is always fixed to $\alpha/p^2$. Even after the elimination of infinitesimal copies, such a result was unchanged. It turns out that this is an exact result, which can be derived by an explicit use of the BRST invariance of the theory. In particular, had the BRST symmetry been violated, such a property would not be valid and the longitudinal piece would pick non-trivial dependencies at higher order loops. 

The two-point function $\langle b^a(x) b^b(y)\rangle$ can be computed by inserting the standard sources coupled to the fields in the partition function and taking two derivatives of it with respect to the source coupled to the field $b$, which is denoted by $J_{(b)}$, i.e.,
\begin{equation}
\langle b^a(x) b^b(y)\rangle = \int [\EuScript{D}\Phi]_{\mathrm{GZ}}\frac{\delta^2}{\delta J^a_{(b)}(x) \delta J^b_{(b)}(y)} \mathrm{e}^{-\Sigma}\,\Bigg|_{J=0}\,,
\label{ELongPart1}
\end{equation}
with
\begin{equation}
\Sigma [\Phi,J] = S_{\mathrm{GZ}}[\Phi] + S_{\mathrm{sources}}[J] + 3V\gamma^4 (N^2-1)\,.
\label{ELongPart2}
\end{equation}
However, since $b^a$ enters in the action at most at quadratic power, it can be integrated out from the path integral leading to the following contribution in the Boltzmann weight of the partition function,
\begin{equation}
\Sigma [\Phi,J] \sim \int_x\Bigg(\frac{(\partial_\mu A^a_\mu)^2}{2\alpha}+\frac{1}{\alpha}J^{a}_{(b)} \partial_\mu A^a_\mu + \frac{J^a_{(b)}J^a_{(b)}}{2\alpha}\Bigg)\,.
\label{ELongPart3}
\end{equation}
The action of the functional derivatives as in \eqref{ELongPart1} leads to
\begin{equation}
\langle b^a(x) b^b(y)\rangle = -\delta^{ab}\frac{\delta (x-y)}{\alpha}+\frac{1}{\alpha^2}\partial^{x}_\mu\partial^{y}_\nu\langle A^{a}_{\mu} (x)A^{b}_{\nu} (y)\rangle\,.
\label{ELongPart4}
\end{equation}
On the other hand, the BRST invariance ensures that
\begin{equation}
\langle b^a(x) b^b(y)\rangle = \langle s\Big(\bar{c}^a(x) b^b(y)\Big)\rangle = 0\,,
\label{ELongPart5}
\end{equation}
and thereby
\begin{equation}
\frac{1}{\alpha}\partial^{x}_\mu\partial^{y}_\nu\langle A^{a}_{\mu} (x)A^{b}_{\nu} (y)\rangle = \delta^{ab}\delta (x-y)\,. 
\label{ELonPart6}
\end{equation}
In the momentum space it translates to
\begin{equation}
p_\mu p_\nu \langle A^{a}_\mu A^b_\nu\rangle (p) = \delta^{ab}\alpha\,.
\label{ELonPart7}
\end{equation}
Decomposing the propagator in its general tensor structure,
\begin{equation}
\langle A^{a}_\mu A^b_\nu\rangle (p)= \delta^{ab}\Bigg[\EuScript{A}\Big(\delta_{\mu\nu}-\frac{p_\mu p_\nu}{p^2}\Big)+\EuScript{B}p_\mu p_\nu + \EuScript{C}\epsilon_{\mu\lambda\nu}p_\lambda\Bigg]\,,
\label{ELonPart8}
\end{equation}
and using \eqref{ELonPart7}, one obtains
\begin{equation}
p_\mu p_\nu\langle A^{a}_\mu A^b_\nu\rangle (p) = \delta^{ab}p^4 \EuScript{B} = \delta^{ab}\alpha\,,
\label{ELonPart9}
\end{equation}
which implies that
\begin{equation}
\EuScript{B} = \frac{\alpha}{p^4}.
\label{ELonPart10}
\end{equation}
As a conclusion, the BRST invariance \eqref{ELongPart5} leads to the conclusion that the longitudinal parity-preserving part of the gauge-field propagator is exact and it does not feel the elimination of infinitesimal Gribov copies as long as it does not affect BRST invariance.
\section{Infrared instabilities and the emergence of condensates}\label{Sect:Condensates}

\subsection{Short overview of the Refined Gribov-Zwanziger origins}

In its original formulation \cite{Zwanziger:1989mf} the Gribov-Zwanziger action, constructed to eliminate infinitesimal Gribov copies in pure Yang-Mills theories, has a striking property: The gluon propagator, in the Landau gauge, exactly vanishes at zero momentum. This has been known as the scaling solution of the gluon propagator in the Landau gauge and it has been obtained by other methods that access the infrared behavior of Yang-Mills theories \cite{Alkofer:2000wg,Maas:2011se,Cyrol:2016tym,Huber:2018ned}. However, more recent gauge-fixed lattice simulations have revealed a finite value at zero momentum for the gauge-field propagator \cite{Cucchieri:2007rg,Cucchieri:2008fc,Cucchieri:2011ig,Maas:2008ri}. In order to circumvent this issue within the Gribov-Zwanziger scenario for pure Yang-Mills theories, it was observed in \cite{Dudal:2007cw,Dudal:2008sp} that non-trivial condensates are formed due to infrared instabilities of the GZ action \cite{Dudal:2011gd}. By taking into account such condensates from the beginning, this leads to the so-called Refined Gribov-Zwanziger (RGZ) action \cite{Dudal:2008sp} and the tree-level gauge field propagator agrees very well with lattice simulations. Moreover, such a framework gives rise to reasonable predictions for the glueball spectrum \cite{Dudal:2010cd}, and the correct sign for the Casimir energy in the MIT bag model \cite{Canfora:2013zna}. In the Landau gauge the RGZ action is written as
\begin{eqnarray}
S^{\mathrm{YM}}_{\mathrm{RGZ}} &=& S^{\mathrm{YM}}_{\mathrm{GZ}}+\frac{m^2}{2}\int_{x^d}A^a_\mu A^a_\mu\nonumber\\
&-& \mu^2\int_{x^d}\Big(\bar{\varphi}^{ab}_\mu\varphi^{ab}_\mu-\bar{\omega}^{ab}_\mu\omega^{ab}_\mu\Big)\,,  
\label{RGZ1}
\end{eqnarray}
where $m^2$ and $\mu^2$ are mass parameters which are fixed by their own gap equations and the localizing auxiliary fields enter as BRST-quartets as in \eqref{Intro7}. The RGZ action in \eqref{RGZ1} is local, renormalizable at all orders in perturbation theory and effectively implements the restriction to the path integral of Yang-Mills theories in $d>2$ to the Gribov region $\Omega$. In $d=2$, the condensates are not formed due to the typical infrared divergences in two dimensions. However, the action \eqref{RGZ1} breaks the BRST invariance explicitly, but in a soft way. As in the case of the GZ action, the BRST invariance is restored by the use of the gauge invariant composite field $A^{h,a}_\mu$. In \cite{Capri:2015ixa,Capri:2015nzw,Pereira:2016fpn,Capri:2018ijg,Capri:2016aqq,Capri:2016gut}, the BRST-invariant RGZ action, in the linear covariant gauges, was formulated and evidence for the formation of the condensates was provided. Its renormalizability was proven in \cite{Capri:2017bfd}.  

In order to provide evidence for the formation of condensates in the GZ modification of YMCS in the linear covariant gauges, we introduce the local composite operators,
\begin{equation}
\EuScript{O}_{A^2} = A^{h,a}_\mu A^{h,a}_\mu\,,    
\label{RGZ2}
\end{equation}
and
\begin{equation}
\EuScript{O}_{\mathrm{aux}} = \bar{\varphi}^{ab}_\mu \varphi^{ab}_\mu - \bar{\omega}^{ab}_\mu \omega^{ab}_\mu\,,    
\label{RGZ3}
\end{equation}
where the auxiliary fields are now the BRST singlets as in \eqref{NPBRSTYMCS7}, to the GZ action modification to the YMCS and compute the vacuum energy in the next subsection.

\subsection{Formation of condensates}

We introduce the operators \eqref{RGZ2} and \eqref{RGZ3} to $S_{\mathrm{GZ}}$ defined in \eqref{NPBRSTYMCS1} by coupling them to the sources $J$ and $\rho$ as follows,
\begin{eqnarray}
\Sigma [\Phi] &=& S_{\mathrm{GZ}}[\Phi]+J\int_x A^{h,a}_\mu A^{h,a}_\mu \nonumber\\
&-& \rho \int_x \Big(\bar{\varphi}^{ab}_\mu \varphi^{ab}_\mu - \bar{\omega}^{ab}_\mu \omega^{ab}_\mu\Big)\,.
\label{RGZ4}    
\end{eqnarray}
At the level of the vacuum energy $\EuScript{E}_v$, the contribution of the condensates is evaluated by,
\begin{equation}
-\frac{\partial\EuScript{E}_v}{\partial J}\Bigg|_{J=\rho=0} = \langle A^{h,a}_\mu A^{h,a}_\mu\rangle\,,    
\label{RGZ5}    
\end{equation}
and 
\begin{equation}
\frac{\partial\EuScript{E}_v}{\partial \rho}\Bigg|_{J=\rho=0} = \langle \bar{\varphi}^{ab}_\mu \varphi^{ab}_\mu - \bar{\omega}^{ab}_\mu \omega^{ab}_\mu\rangle\,.    
\label{RGZ6}    
\end{equation}
The vacuum energy at one-loop, in the presence of $J$ and $\rho$, is
\begin{equation}
\EuScript{E}_v = \frac{1}{2V}\mathrm{Tr}\,\mathrm{ln}\,, \Delta^{ab}_{\mu\nu}    
\label{RGZ7}    
\end{equation}
with
\begin{eqnarray}
\Delta^{ab}_{\mu\nu} &=& \delta^{ab}\,\Bigg[\delta_{\mu\nu} - \Bigg(1-\frac{1}{\alpha}\Bigg) + M \epsilon_{\mu\lambda\nu}p_\lambda + 2\,\EuScript{P}^{\mathrm{T}}_{\mu\nu}(p)\nonumber\\
&\times&\Bigg(J+\frac{g^2\gamma^4 N}{p^2+\rho}\Bigg)\Bigg]\,.
\label{RGZ8}
\end{eqnarray}
Therefore,
\begin{equation}
\langle A^{h,a}_\mu A^{h,a}_\mu\rangle = 2 (N^2-1)(\EuScript{I}_{1}(\gamma,M)+\EuScript{I}_{2}(\gamma,M))\,,
\label{RGZ9}
\end{equation}
with
\begin{equation}
\EuScript{I}_{1}(\gamma,M)=\int_p \frac{2g^2 \gamma^4 N (p^4+2g^2\gamma^4 N)}{p^2((p^4+2g^2\gamma^4 N)^2+M^2 p^6)}\,,
\label{RGZ10}    
\end{equation}
and
\begin{equation}
\EuScript{I}_2 (\gamma, M) = \int_p\frac{M^2 p^6}{p^2((p^4+2g^2\gamma^4 N)^2+M^2 p^6)}\,.    
\label{RGZ11}
\end{equation}
As for the other condensate $\EuScript{O}_{\mathrm{aux}}$ one obtains
\begin{equation}
\langle \bar{\varphi}^{ab}_\mu \varphi^{ab}_\mu - \bar{\omega}^{ab}_\mu \omega^{ab}_\mu\rangle = -(N^2-1)\,\EuScript{I}_1 (\gamma,M)\,.
\label{RGZ12}    
\end{equation}
From equations \eqref{RGZ10} and \eqref{RGZ11}, we see that the integrals that enter the evaluation of the condensates are convergent. Moreover, $\EuScript{I}_1 (\gamma,M)$ is proportional to the Gribov parameter $\gamma$. Hence, as long as the elimination of infinitesimal Gribov copies is employed, the condensates are generated, see \eqref{RGZ9} and \eqref{RGZ12}. Furthermore, the integral $\EuScript{I}_2 (\gamma,M)$, which enters the condensate $\langle A^{h,a}_\mu A^{h,a}_\mu\rangle$, is such that if $\gamma\to 0$, it still leads to a non-vanishing condensate due to the presence of the topological mass parameter $M$. If $M\to 0$, then this integral also vanishes and, at one-loop, there is no evidence for the formation of condensates.

Due to the previous discussions, we can write the Refined Gribov-Zwanziger modification to the YMCS theory in the linear covariant gauges,
\begin{eqnarray}
S_{\mathrm{RGZ}}[\phi] &=& S_{\mathrm{GZ}}[\phi] + \frac{m^2}{2}\int_x A^{h,a}_\mu A^{h,a}_\mu \nonumber\\
&-& \mu^2\int_x \Big(\bar{\varphi}^{ab}_\mu \varphi^{ab}_\mu - \bar{\omega}^{ab}_\mu \omega^{ab}_\mu\Big)\,,
\label{RGZ13}    
\end{eqnarray}
with $S_{\mathrm{GZ}}$ given by \eqref{NPBRSTYMCS1}. Action \eqref{RGZ13} is invariant under \eqref{NPBRSTYMCS7} and the mass parameters $m^2$ and $\mu^2$ are determined by their own gap equation. Moreover, since they are not coupled do BRST-exact terms, they can enter correlation functions of gauge-invariant operators. \begin{eqnarray}
\langle A^a_\mu (p) A^b_\nu (-p)\rangle &=& \delta^{ab}\mathcal{D}(p^2)\,\Bigg[\EuScript{P}^{\mathrm{T}}_{\mu\nu}(p) - \mathcal{G}(p^2)\epsilon_{\mu\lambda\nu}p_\lambda\Bigg]\nonumber\\
&+&\delta^{ab}\frac{\alpha}{p^2}\frac{p_\mu p_\nu}{p^2}\,,
\label{RGZ14}
\end{eqnarray}
with
\begin{widetext}
\begin{equation}
\mathcal{D}(p^2) = \frac{(p^2+\mu^2)[(p^2+m^2)(p^2+\mu^2)+2g^2\gamma^4 N]}{[(p^2+m^2)(p^2+\mu^2)+2g^2\gamma^4 N]^2 + M^2p^2 (p^2+\mu^2)^2}\,,
\label{RGZ15}    
\end{equation}
\end{widetext}
and
\begin{equation}
\mathcal{G}(p^2) = \frac{M(p^2+\mu^2)}{(p^2+m^2)(p^2+\mu^2)+2g^2\gamma^4 N}\,.    
\label{RGZ16}
\end{equation}
Once again, one sees that due to the BRST-invariance of \eqref{RGZ13} the parity-preserving longitudinal part of the gauge-field propagator is the same as the one of YMCS theory in linear covariant gauges without the elimination of infinitesimal copies. This is a powerful property that follows from the BRST invariance of the entire construction. The inclusion of condensates will affect the pole structure of the gauge field propagator. However, it is beyond the scope of the present paper to solve the gap equations, which determines the corresponding gap equations associated to each mass parameter. Therefore, we have several parameters which would be treated as free parameters making the analysis cumbersome and not so informative. Hence, we leave this for future work. Non-perturbative propagators in linear covariant gauges were reported also by different methods in pure Yang-Mills theories, see \cite{Huber:2015ria,Aguilar:2015nqa,Aguilar:2016ock,Napetschnig:2021ria}.

\section{Conclusions}\label{Sect:CONCLUSIONS}

Three dimensional non-Abelian gauge theories can combine the non-trivial dynamics of Yang-Mills theories with the Chern-Simons action. The YMCS features infinitesimal gauge invariance when the topological Chern-Simons mass is unconstrained and, therefore, requires a gauge-fixing procedure for practical calculations of quantum effects in the continuum. The existence of Gribov copies in the Faddeev-Popov method and their manifestation in the low-energy regime of the theory suggest that the gauge-fixing prescription is amended in the infrared. For such a model, this was explored in \cite{Canfora:2013zza,Gomez:2015aba,Ferreira:2020kqv} in the Landau and maximal Abelian gauges. In this work, we extended the elimination of infinitesimal Gribov copies in the linear covariant gauges. While in the Landau and maximal Abelian gauges the elimination breaks BRST invariance, in this work we show how to restore it for color- and Lorentz-covariant linear gauges. In the maximal Abelian gauge, it is also possible to construct a BRST-invariant action free of infinitesimal Gribov copies as discussed in \cite{Capri:2015pfa}. We leave this out of the present work since our main focus was to explore the role of the gauge parameter $\alpha$ in linear covariant gauges. We have proved that the presence of the Chern-Simons term as well as the BRST-invariant Gribov-Zwanziger modification to the YMCS do not affect the longitudinal parity-preserving component of the gauge-field propagator as in standard Yang-Mills theory. This is a profound consequence of BRST invariance within a linear gauge condition.

Next to that, we have explored whether the elimination of the Gribov copies, as in pure Yang-Mills theories, could generate further infrared instabilities such as the formation of condensates. As shown by an explicit one-loop calculation, BRST-invariant condensates $\langle A^{h,a}_\mu A^{h,a}_\mu\rangle$ and $\langle \bar{\varphi}^{ab}_\mu {\varphi}^{ab}_\mu - \bar{\omega}^{ab}_\mu {\omega}^{ab}_\mu\rangle$ are generated and are proportional to the Gribov parameter, which is characteristic of the elimination of Gribov copies. This gives rise to the refined Gribov-Zwanziger modification to the YMCS. The tree-level gauge-field propagator was computed and, once again, thanks to the BRST-invariance of the formulation, the longitudinal component of the parity-preserving gauge-field propagator is the same as its tree-level value in standard Yang-Mills theories and, moreover, exact to all orders in perturbation theory.

As for new perspectives to the present line of research, it is essential to establish the renormalization properties of the present model. Since the BRST-invariant RGZ modification to the YMCS in linear covariant gauges involves the non-polynomial field $A^h_\mu$, this is a subtle issue. However, in the same lines as in \cite{Capri:2017bfd}, it can be investigated by means of the algebraic renormalization framework \cite{Piguet:1995er}. Another interesting avenue to be investigated is the coupling of matter fields and how their dynamics can impact the pole structure of the non-perturbative propagators presented herein, see, e.g., \cite{Gomez:2015aba} and, also, how the non-perturbative effects of the restriction to the Gribov region can affect the propagators of matter fields. Those issues are left for future work.
\\

\section*{Acknowledgments}

ADP acknowledges CNPq under the grant PQ-2 (309781/2019-1), FAPERJ under the “Jovem Cientista do Nosso Estado” program (E26/202.800/2019), and NWO under the VENI Grant (VI.Veni.192.109) for financial support. IFJ acknowledges CAPES for the financial support under the project grant $88887.357904/2019-00$.

\appendix

\begin{widetext}
\section{Construction of $A^h$}\label{Appendix:AhConstruction}
In this Appendix, we collect properties of the gauge invariant $A^h_\mu$ field. We begin with the definition of the functional $f_{A}[u]$ written as
\begin{equation}
f_{A}[u]\equiv \mathrm{Tr\,}\int \mathrm{d}^dx\,A^{u}_{\mu}A^{u}_{\mu}= \mathrm{Tr\,}\int \mathrm{d}^dx\,\left(u^{\dagger}A_{\mu}u+\frac{i}{g}u^{\dagger}\partial_{\mu}u\right)\left(u^{\dagger}A_{\mu}u+\frac{i}{g}u^{\dagger}\partial_{\mu}u\right)\,.
\label{ah1}
\end{equation}
For a given gauge field configuration $A_{\mu}$, $f_{A}[u]$ is a functional over its gauge orbit. A minimum $f_{A}[h]$ is attained when
\begin{eqnarray}
\delta f_{A}[u]\big|_{u=h} &=& 0\,,\nonumber\\
\delta^2 f_{A}[u]\big|_{u=h} &>& 0\,,
\label{ah2}
\end{eqnarray}
and it is an absolute minimum if
\begin{equation}
f_{A}[h] \leq f_{A}[u]\,,\,\,\, \forall u\,\,\in\,\, \mathcal{U}\,,
\label{ah3}
\end{equation}
where $\mathcal{U}$ is the set of local gauge transformations. With the absolute minimum $f_{A}[h]$ at our disposal, it is possible to define a gauge invariant quantity by
\begin{equation}
\mathcal{A}^{2}_{\mathrm{min}}=\underset{\left\{u\right\}}{\mathrm{min}}\,\mathrm{Tr\,}\int d^4x\,A^{u}_{\mu}A^{u}_{\mu}=f_{A}[h]\,.
\label{ah4}
\end{equation}
Searching for absolute minimum is an extremely challenging task. However, one can collect at least the relative minimum by demanding conditions (\ref{ah2}). This can be achieved by an expansion on the coupling $g$. We define
\begin{equation}
v=h\mathrm{e}^{ig\omega}\equiv h\mathrm{e}^{ig\omega^{A}T^A}\,,
\label{ah5}
\end{equation} 
with $T^A$ the $SU(N)$ generators and $\omega^A$ a small parameter. Due to this assumption, we retain terms up to $\omega^2$, which is enough for our purposes. By definition,
\begin{eqnarray}
A^{v}_{\mu}&=& v^{\dagger}A_{\mu}v\frac{i}{g}v^{\dagger}\partial_{\mu}v\nonumber\\
&=&\mathrm{e}^{-ig\omega}h^{\dagger}A_{\mu}h\mathrm{e}^{ig\omega}+\frac{i}{g}\mathrm{e}^{-ig\omega}h^{\dagger}(\partial_{\mu}h)\mathrm{e}^{ig\omega}+\frac{i}{g}\mathrm{e}^{-ig\omega}\partial_{\mu}\mathrm{e}^{ig\omega}\nonumber\\
&=&\mathrm{e}^{-ig\omega}A^{h}_{\mu}\mathrm{e}^{ig\omega}+\frac{i}{g}\mathrm{e}^{-ig\omega}\partial_{\mu}\mathrm{e}^{ig\omega}\,,
\label{ah6}
\end{eqnarray}
where we the definition of $A^h_{\mu}$ was used and $h^{\dagger}h=1$. Expanding eq.(\ref{ah6}) up to quadratic order in $\omega$, we obtain
\begin{eqnarray}
A^{v}_{\mu}&=&\left(1-ig\omega-\frac{g^2}{2}\omega^2+\mathcal{O}(\omega^3)\right)A^{h}_{\mu}\left(1+ig\omega-\frac{g^2}{2}\omega^2+\mathcal{O}(\omega^3)\right)+\frac{i}{g}\left(1-ig\omega-\frac{g^2}{2}\omega^2\right)\partial_{\mu}\left(1+ig\omega-\frac{g^2}{2}\omega^2\right)+\mathcal{O}(\omega^3)\nonumber\\
&=& A^h_{\mu}+igA^{h}_{\mu}\omega-\frac{g^{2}}{2}A^{h}_{\mu}\omega^2-ig\omega A^{h}_{\mu}+g^2\omega A^{h}_{\mu}\omega-\frac{g^2}{2}\omega^2 A^{h}_{\mu}+\frac{i}{g}\left(ig\partial_{\mu}\omega\phantom{\frac{1}{2}}-\frac{g^2}{2}(\partial_{\mu}\omega)\omega-\frac{g^2}{2}\omega\partial_{\mu}\omega+g^2\omega\partial_{\mu}\omega\right)\nonumber\\
&+&\mathcal{O}(\omega^3)\,.
\label{ah7}
\end{eqnarray}
After a few simple manipulations,  eq.(\ref{ah7}) becomes
\begin{equation}
A^{v}_{\mu}=A^{h}_{\mu}-\partial_{\mu}\omega+\frac{ig}{2}[\omega,\partial_{\mu}\omega]+ig[A^{h}_{\mu},\omega]+\frac{g^2}{2}[[\omega,A^{h}_{\mu}],\omega]+\mathcal{O}(\omega^3)\,.
\label{ah8}
\end{equation}
Now, we explicitly compute $f_{A}[v]$,
\begin{eqnarray}
f_{A}[v]&=&\mathrm{Tr\,}\int \mathrm{d}^4x\,A^{v}_{\mu}A^{v}_{\mu}= \mathrm{Tr\,}\int \mathrm{d}^4x\,\left(A^{h}_{\mu}-\partial_{\mu}\omega+\frac{ig}{2}[\omega,\partial_{\mu}\omega]+ig[A^{h}_{\mu},\omega]+\frac{g^2}{2}[[\omega,A^{h}_{\mu}],\omega]+\mathcal{O}(\omega^3)\right)\nonumber\\
&\times&\left(A^{h}_{\mu}-\partial_{\mu}\omega+\frac{ig}{2}[\omega,\partial_{\mu}\omega]+ig[A^{h}_{\mu},\omega]+\frac{g^2}{2}[[\omega,A^{h}_{\mu}],\omega]+\mathcal{O}(\omega^3)\right)
\label{ah9}
\end{eqnarray}
which implies
\begin{equation}
f_{A}[v]=
f_{A}[h]+2\,\mathrm{Tr\,}\int \mathrm{d}^4x\,\omega(\partial_{\mu}A^{h}_{\mu})-\mathrm{Tr\,}\int \mathrm{d}^4x\,\omega\partial_{\mu}D_{\mu}(A^h)\omega+\mathcal{O}(\omega^3)\,.
\label{ah11}
\end{equation}
Condition (\ref{ah2}) is automatically satisfied for
\begin{eqnarray}
\partial_{\mu}A^{h}_{\mu}&=& 0 \nonumber\\
-\partial_{\mu}D_{\mu}(A^h)&>& 0\,.
\label{ah12}
\end{eqnarray}
Using the transversality condition $\partial_{\mu}A^{h}_{\mu} = 0$, one can solve $h=h(A)$ as a power series in $A_{\mu}$. As a result, one writes $A^h_{\mu}=A^h_{\mu}(A)$. Starting from the definition of $A^h_{\mu}$, eq.\eqref{NPBRSTYMCS2} and considering that 
\begin{equation}
h=\mathrm{e}^{ig\phi^AT^A}\equiv\mathrm{e}^{ig\phi}=1+ig\phi-\frac{g^2}{2}\phi^2+\mathcal{O}(\phi^3)\,,
\label{ah14}
\end{equation}
leads to
\begin{equation}
A^{h}_{\mu}=A_{\mu}+ig[A_{\mu},\phi]-\frac{g^2}{2}A_{\mu}\phi^2+g^2\phi A_{\mu}\phi-\frac{g^2}{2}\phi^2 A_{\mu}-\partial_{\mu}\phi+\frac{ig}{2}[\phi,\partial_{\mu}\phi]+\mathcal{O}(\phi^3)\,.
\label{ah16}
\end{equation}
Imposing the transversality of $A^{h}_{\mu}$ on (\ref{ah16}), allows for a solution of $\phi$ in terms of $A_\mu$,
\begin{equation}
\phi=\frac{1}{\partial^2}\partial A + \frac{ig}{2}\frac{1}{\partial^2}\left[\partial A,\frac{1}{\partial^2}\partial A\right]+ig\frac{1}{\partial^2}\left[A_{\alpha},\frac{\partial_{\alpha}}{\partial^2}\partial A\right]+\mathcal{O}(A^3)\,.
\label{ah18}
\end{equation}
Substituting eq.(\ref{ah18}) in (\ref{ah16}), one gets an explicit expression for $A^h_{\mu}$ as a power series of $A_{\mu}$,
\begin{eqnarray}
A^{h}_{\mu}&=&A_{\mu}-\partial_{\mu}\frac{1}{\partial^2}\partial A+ig\left[A_{\mu},\frac{1}{\partial^2}\partial A\right]-ig\frac{1}{\partial^2}\partial_{\mu}\left[A_{\alpha},\partial_{\alpha}\frac{1}{\partial^2}\partial A\right]+\frac{ig}{2}\frac{1}{\partial^2}\partial_{\mu}\left[\frac{1}{\partial^2}\partial A,\partial A\right]\nonumber\\
&+&\frac{ig}{2}\left[\frac{1}{\partial^2}\partial A,\partial_{\mu}\frac{1}{\partial^2}\partial A\right] + \mathcal{O}(A^3)\,.
\label{ah19}
\end{eqnarray}
This highly non-local structure is gauge-invariant order by order in $g$.
\end{widetext}
\bibliography{refsymcsLCG}

\end{document}